\documentclass{ws-procs975x65}

\def\lsim{\lower.5ex\hbox{$\; \buildrel < \over \sim \;$}}
\def\gsim{\lower.5ex\hbox{$\; \buildrel > \over \sim \;$}}

\begin{document}
\vskip -2.0cm
\title{QUASI PERIODIC OSCILLATIONS DUE TO AXISYMMETRIC AND NON-AXISYMMETRIC SHOCK OSCILLATIONS IN BLACK HOLE ACCRETION}

\author{SANDIP K. CHAKRABARTI$^{1,2}$, D. DEBNATH$^2$, P.S. PAL$^2$, A. NANDI$^2$, R. SARKAR$^2$, M.M. SAMANTA$^2$,
P.J. WIITA$^3$, H. GHOSH$^{1}$ and D. SOM$^{1}$}

\address{$^1$S. N. Bose National Centre for Basic Sciences, JD-Block, Sector III,\\
  Salt Lake, Kolkata, 700 098, India. E-mail:chakraba@bose.res.in\\
  $^2$ Centre for Space Physics, Chalantika 43, Garia Station Rd., Kolkata 700084, India.\\
$^3$ Department of Physics \& Astronomy, Georgia State University, PO Box 4106, Atlanta, GA 30302-4106, USA\\}

\begin{abstract}

Quasi-Periodic Oscillations (QPOs) are very puzzling since they remain totally unexplained by 
popular earlier models of accretion disks. The significant rms value in power density spectrum implies 
that the oscillation involves in the dynamical and non-linear variation of certain region of the 
accretion disk itself. The nature of the energy dependence implies that the region 
which produces Comptonized hard tail is also responsible for QPOs. Similarly, the occurrences of the
QPOs are strongly related to the jet formation and the spectral states. These features are
the natural consequences of the advective disk paradigm that we are advocating.
In the mid 90s, some of the present authors first pointed out that the QPOs in all possible types of black holes
may be simply due to the oscillations of the CENBOL, the CENtrifugal pressure supported BOundary Layer
which is formed in the sub-Keplerian flows around a black hole. This CENBOL could 
be axi-symmetric as well as non-axisymteric in nature since its boundary, 
namely, the centrifugally driven shocks could be axi-symmetric or non-axisymmetric.
In addition, we pointed out that the transition radius where the flow becomes Keplerian 
to sub-Keplerian, as well as the location of the inner sonic point can also oscillate and produce 
the QPOs. Since the shock locations are functions of the 
specific angular momentum ($\lambda$) and specific energy (${\cal E}$) of the flow, our model 
naturally predicts that the QPO frequency should vary with mass, spin, $\lambda$ and ${\cal E}$.
The QPO frequencies with specific ratios, such as, 2:3 must be due to non-axisymmetric
effects when the shock switches between the two-armed and the three-armed spirals. We also
discuss the possible effects that the disk inclination might have with the line of sight.
\end{abstract}


\bodymatter

\section{Introduction}

Quasi-Periodic Oscillations (QPOs) of radiations emitted from the accretion disks of
black hole candidates remain totally unexplained by the `standard' 
model of accretion disk (Shakura \& Sunyaev, 1973) or Advection Dominated Accretion Disk 
(Esin et al. 1998) or any of their variations. The oscillations manifest themselves so powerfully 
that they leave little doubt that they are not merely due to vibrations of the disks 
or jets. Rather, regions of the disks are dynamically oscillating and causing the
modulation in radiation intensity. Moreover, while the thickness of the region determines the `Q' value of the
QPO, the size of that region determines the frequency itself. The movement of that region
causes drifting of the frequency. Thus a gradual or rapid increase of frequency means 
that the region moves towards the black hole, while the decrease of frequency means that the region 
is drifting away. Another interesting and quite general property of QPOs is 
that with the increase in luminosity, the frequency goes up, as if the region moves
closer to the black hole. Also, the higher energy photons appear to have sharper QPOs. 
In very soft states QPOs disappear.

What is this `region' which causes QPOs in black hole and neutron star candidates? Guesses and 
counter-guesses are rampant ranging from blobby disks, gravitational lensing,
some instability at inner stable orbit to elliptical precessing trajectories in the
disks. In reality, the QPOs are the most natural manifestation of a non-steady sub-Keplerian
flow in which cooling through radiative effects or dynamical cooling through jets 
is important. We have argued for over a decade now that 
QPOs are due to oscillations of the CENBOL, the CENtrifugal pressure supported BOundary Layer
which is bounded by a shock and the sonic point. 
With every new observation this assertion is vindicated even further. The oscillation
frequency is roughly the infall time of matter in the CENBOL. With increase
in luminosity,the post-shock region cools down and the shock moves closer and the 
frequency is also increased.  This review is to show the present status of the QPOs 
in black hole candidates, especially emphasizing the occurrences of QPO frequencies in 2:3 ratio.

Even though there are popular `models' which can `explain' the
`frequencies' of QPOs by vibrations of the disk, or Keplerian frequency at the
inner stable orbit, or trapped oscillations, etc. these models can be numerology
at the best. By definition, any perturbation would oscillate with a frequency close
or equal to the Keplerian frequency, whether it is perturbed vertically or horizontally
or any other way. One can always 
identify that with QPO frequency. So, explaining the frequency is not an issue. 
The major issue is to answer the question: does any solution (not a model) actually reproduce 
the power density spectra, break frequency, the power at the QPO frequency, 
even the multitude of QPOs observed on the same day, {\it and} explain the 
variation of QPO frequency with luminosity etc. ?
We are convinced that since we stick to the proper (time dependent) solutions of the governing equations,
our explanation is the most natural and universal. If there are deficiencies in terms of details,
the problem lies not in our approach, but in the non-inclusion of all the physical 
processes in the governing equations.

\section{QPOs from Axisymmetric Oscillations}

To some, the inner stable circular orbit (ISCO)
has {\it always} some thing to do with the quasi-periodic oscillations (QPOs).
This is totally wrong since for a given black hole candidate (i.e., for a given ISCO location, and thus
frequency of the Keplerian orbit), there can be a wide variation in QPO frequency,
often ranging from a few mHz to a few Hz (see, works of Remillard et al. 1999 and references therein
and Chakrabarti \& Manickam, 2000; Chakrabarti et al. 2006). Very often, several QPO
frequencies are detected simultaneously. As stated in the introduction, since
QPOs have significant power (a few percent of rms value) it
is not a `vibration' (i.e., disko-seismological effects) of something.
Rather, they are due to oscillations of various regions as a whole. In a two component advective paradigm
there are several length scales which may participate in `natural' oscillations.
These are: (a) $r=r_{Kep}$ which is the inner edge of the Keplerian disk
which need not be ISCO. $r_{Kep}$ could extend to even a few hundred $r_g$,
the Schwarzschild radius. (b) $r=r_{s}$ is the mean location of the shock.
This can vary from $\sim 5 r_g$ to $100 r_g$ depending on specific energy
and specific angular momentum. For a Kerr black hole the lower limit
could be even closer to the black hole. (c) $r=r_{in}$, the
location of the inner sonic point which varies between the marginally stable and the
marginally bound orbits, i.e., between $2$ to $3$ Schwarzschild radii
for a Schwarzschild black hole. For Kerr black holes, both of these orbits are closer to the
black hole and thus the frequency can rise. In reality, it is the inner strong
shock $r_{in}\lsim r_{ws} \lsim 10 $ with a typical location of $r\sim 7$
just behind $r_{in}$ is more important (Samanta, Chakrabarti \& Ryu, 2007)
for the purpose of radiative transfer (see Figs. 8-10 below). (d) $r=r_{sr}$, the
size of the sonic radius of the outflow or jet which comes out of the
CENBOL. This is where the jet becomes supersonic and is usually a few times the shock location (Chakrabarti,
1999; Chakrabarti \& Nandi, 2000). The outflow till $r_{sr}$ being sub-sonic,
it is denser and can be cooled by the soft photons from the disk,
especially from the pre-shock flow ($r>r_s$). The cold outflow then falls back and is recycled
through the accretion disks again. This process may take place at an interval of a few seconds
and cause the low frequency QPO.
                                                                                                                           
The four length scales mentioned here scale with the
mass of the black holes. They also correspond to four time scales and hence four types of QPOs
whose frequencies scale as the inverse of the mass. The QPO frequency $\nu$
roughly varies as the inverse of the infall time $t_{QPO}$ from one of the first three
types of the oscillating scales $r$ to the horizon, i.e., $t_{QPO} \sim \nu^{-1} \sim R r^{3/2}$ $2GM/c^3$s,
where $R$ is the shock strength ($R=1$ for $r=r_{Kep}$) which could be $\sim 7$ for a
strong shock and $M$ is the mass of the black hole, be in of stellar mass or super-massive. For a
$M=10^8M_\odot$ Schwarzschild black hole, for instance, the QPO frequency will be
(a) $0.027  < \nu_a < 27.5 \mu$Hz for $3 < r < 300$, (b) $0.14 < \nu_b <13\mu$Hz for $R=7$,
and $5 < r < 100$, (c) $\nu_c \sim 7.7\mu$Hz for typical values ($R=7$, $r_{ws}=7$).
The fourth type of oscillation depends on the time in which the volume $r_{sr}^3$ is filled in
till the optical depth $\tau \sim 1$ (Chakrabarti \& Manickam, 2000). 
This oscillations is of low frequency ($\sim$ mHz) and brings the spectrum from on to off states
in quick succession (few minutes) due to local recycling of matter. Chakrabarti \& Manickam (2000)
showed that the duration (recycling time $t_{rec}$) of the QPO, depends on the QPO frequency $\nu_{b}$
(in the off-state) itself through the relation $t_{rec}=\nu_{d}^{-1} \propto \nu_b^{-2}$. So, $\nu_d$
is not independent and depends on $\nu_b$.

Theoretical works (Chakrabarti, 1989, 1990) on axisymmetric shocks in accretion disks
show that  standing shocks are possible  for a large region of the parameter space. This
conclusion remains practically the same even when moderate viscosities are 
present (Chakrabarti \& Das, 2004). These solutions were found to be correct when
numerical simulations were carried out. The first successful numerical simulation of time dependent 
shock solution was carried out in 1994 (Molteni, Sponholz and Chakrabarti, 1996,
hereafter MSC96) and it was immediately realized that QPOs are due to shock 
oscillations. Early numerical simulations
(Chakrabarti \& Molteni, 1993; Molteni, Lanzafame \& Chakrabarti, 1994) clearly showed that the centrifugal
barrier causes axisymmetric standing shocks to form around black holes and further theoretical
(Chakrabarti, 1990; Chakrabarti \& Das 2004) and numerical simulations (Lanzafame, Molteni \& Chakrabarti, 1998)
indicated that when the viscosity is high enough, the standing shock disappears.
Figures 1(a-b) show two stages of the accretion flow configuration at half-cycle intervals
in the simulation of MSC96. The CENBOL region forms and collapses quasi-periodically.
Here, a power-law cooling $\propto \rho^2T^\alpha$ was used, $\alpha=0.5$ corresponds to
bremsstrahlung. Physically, cooling reduces the post-shock pressure and moves the shock
inward at a steady location. But when it is perturbed, say pushed backward, the post-shock temperature
goes up as the relative motion between the pre-shock and post-shock rises. This causes excess cooling
and the shock collapses toward the black hole only to be bounced back due to centrifugal force.
The calculation of infall time is not easy since the presence of
turbulence causes the flow to deviate from falling freely. On an average, it is seen that a constant velocity
in the post-shock region, at least up to the inner sonic point, is a better guess (Chakrabarti \& Manickam, 2000).
Another way is to assume the infall time from $r$ to be $\sim R r/v$, where, $R$ is the shock strength (ratio
of the post-shock and pre-shock densities) and $v$ is the infall velocity $\sim 1/r^{1/2}$.

\def\figsubcap#1{\par\noindent\centering\footnotesize(#1)}
\begin{figure}[b]%
\begin{center}
\vskip 0.0 in
\hskip -4.0 in
  \parbox{1.0in}{\epsfig{figure=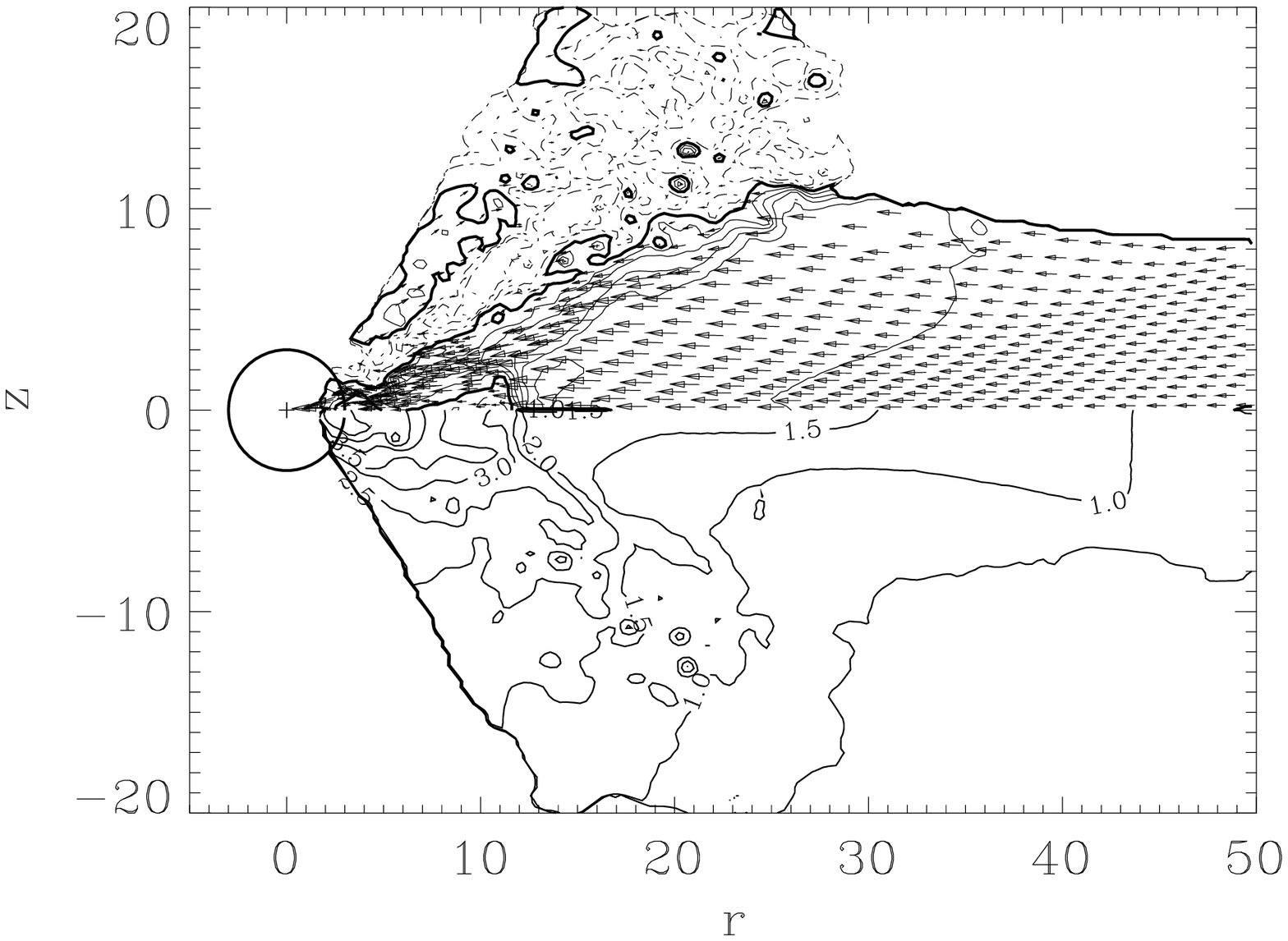,width=4.5in}}
\end{center}
\begin{center}
\hskip -4.0 in
  \parbox{1.0in}{\epsfig{figure=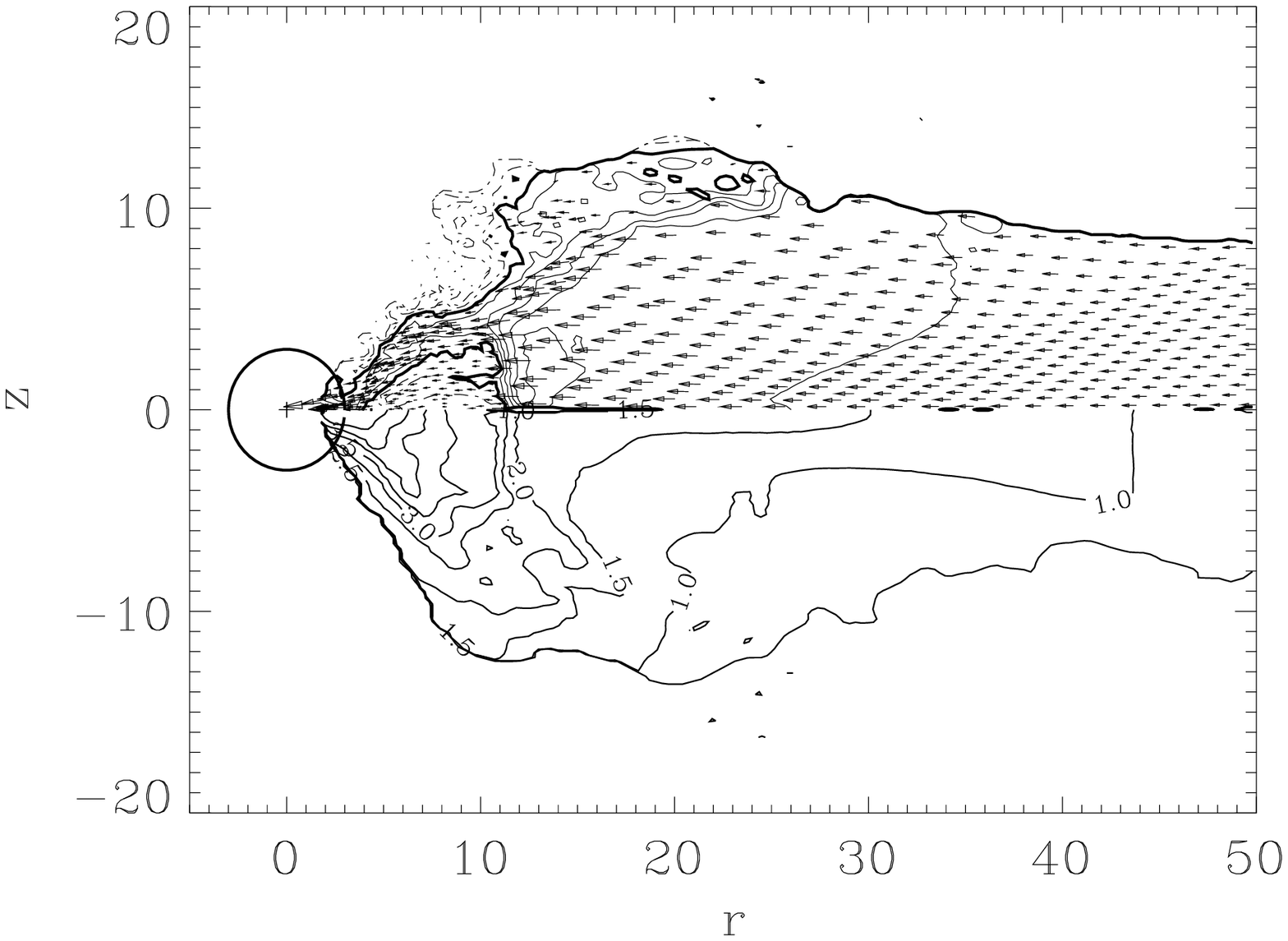,width=4.5in}}
\caption{ Flow topologies at two phases of the shock oscillations
in simulations in which only the radial oscillation of the shocks are allowed (from MSC96).
Power law cooling has been used. Shock oscillates due to the matching of the infall time 
and the cooling time.}
\end{center}
\end{figure}

\def\figsubcap#1{\par\noindent\centering\footnotesize(#1)}
\begin{figure}[b]%
\begin{center}
\vskip 0.0 in
\hskip -3.0 in
  \parbox{1.0in}{\epsfig{figure=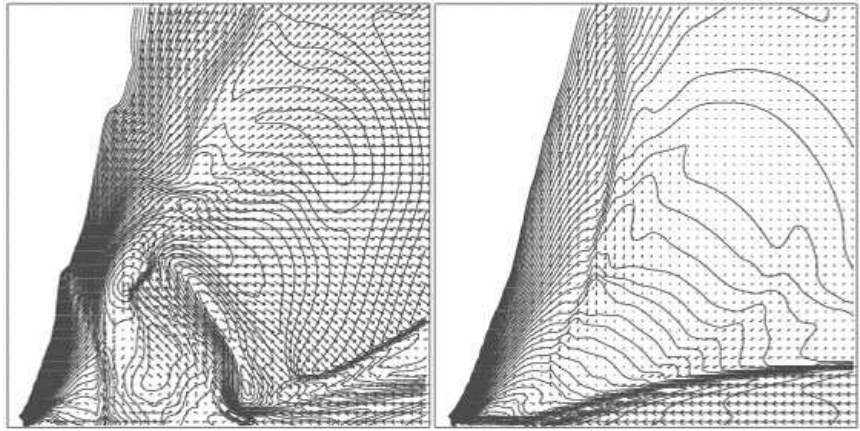,width=4.5in}}
\caption{Comparison of the flow topologies at two phases of the shock oscillations
in simulations in which only the radial oscillation of the shocks are allowed (from RCM97).
Here no cooling was used but the energy and angular momentum were so chosen that
no standing shock is formed.}
\end{center}
\end{figure}

\begin{figure}
\begin{center}
\vskip -1.0cm
\hskip -2.0 in
\parbox{1.0in}{\epsfig{figure=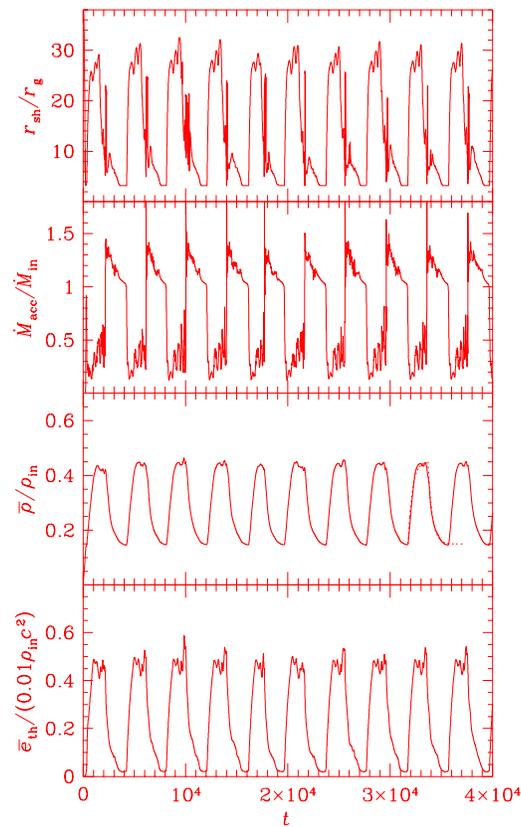,width=3.5in}}
\caption{Time dependence of the average shock location, accretion rate, average 
density and average thermal energy during sustained oscillation of an advective, 
sub-Keplerian disk (from RCM97).}
\end{center}
\end{figure}
                                                                                                                             
\begin{figure}
\begin{center}
\vskip -5.0cm
\hskip -8.0cm
\parbox{1.0in}{\epsfig{figure=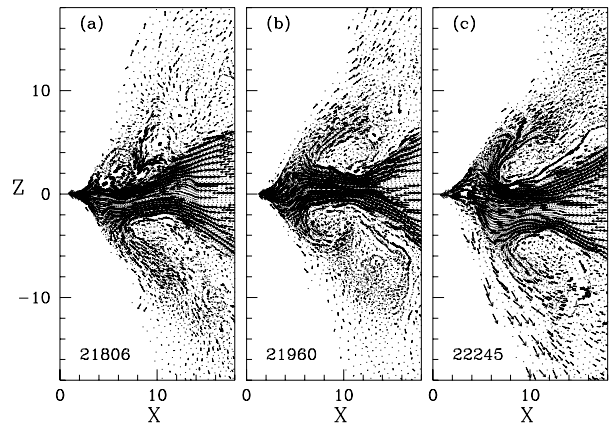,width=4.0in}}
\caption{Configuration of the advective flow with cooling effects at three different
phases of oscillation when both the radial and the vertical motions of the shocks are allowed
(i.e., no symmetry is imposed). The modulation of X-rays due to this oscillation causes
quasi-period oscillations in black hole candidates (from CAM04).}
\end{center}
\end{figure}
                                                                                                                             
\begin{figure}
\begin{center}
\vskip -0.0cm
\hskip -6.0cm
\parbox{1.0in}{\epsfig{figure=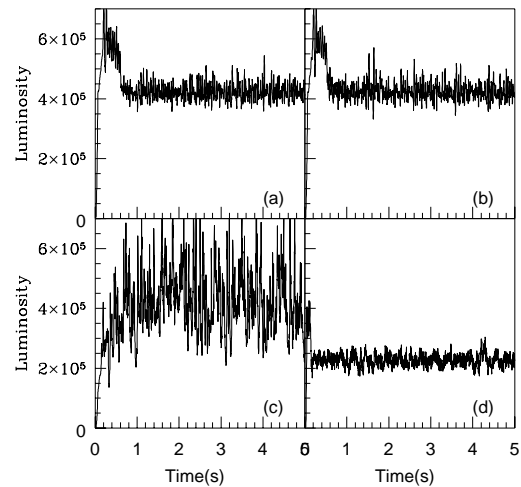,width=4.0in}}
\caption{Variation of emitted luminosity (in arbitrary units) with time from four numerical simulations
in increasing order of accretion rate (from CAM04). Dimensionless accretion rates (of the sub-Keplerian
flow) when the cooling is normalized to Compton cooling are (a) 0.05, (b) 0.08, (c) 0.39 and (d) 0.85.}
\vskip -2.0cm
\end{center}
\end{figure}
                                                                                                                             
\begin{figure}
\begin{center}
\vskip -2.0cm
\hskip -5.5cm
\parbox{1.0in}{\epsfig{figure=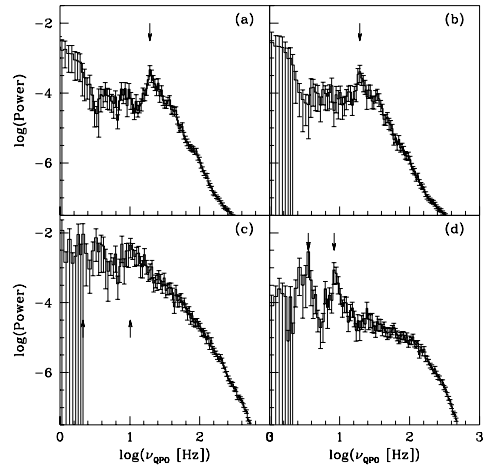,width=4.0in}}
\vskip -2.0cm
\caption{Variation of Power Density Spectra (PDS) for the cases presented in Fig. 5.
The QPO frequencies are (a) 19.34 Hz, (b) 19.45 Hz, (c) 10.2 Hz and 2.67 Hz and (d) 8.32 Hz and 3.58 Hz (from CAM04).
QPOs are marked with arrows.}
\end{center}
\end{figure}
                                                                                                                             
\begin{figure}
\begin{center}
\vskip -0.0cm
\hskip -8.5cm
\parbox{1.0in}{\epsfig{figure=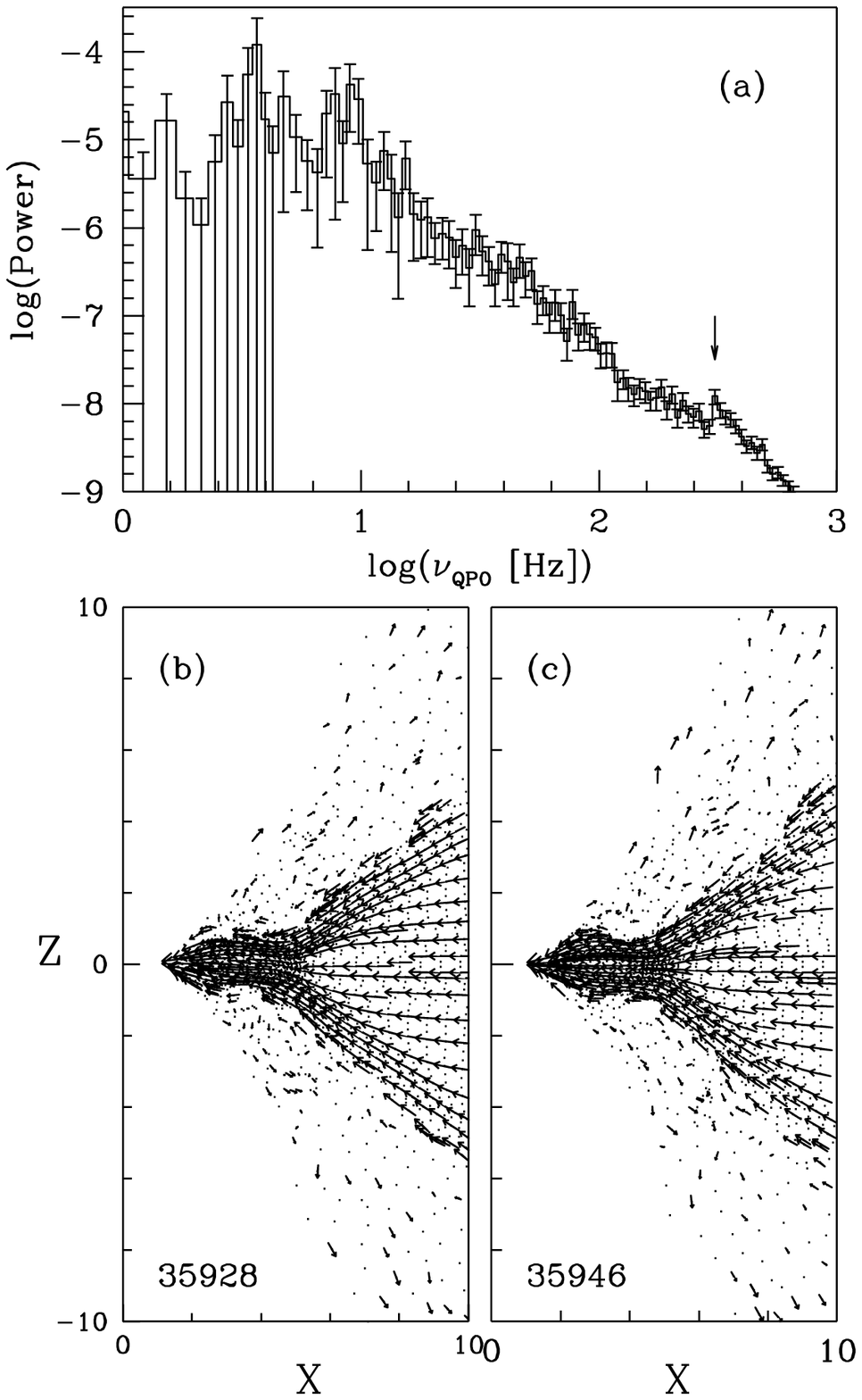,width=4.0in}}
\caption{The simulation results reproduce high frequency QPOs at $\nu\sim 330$Hz as seen in
the PDS (a). The flow configurations in (b) show that this is due to the oscillation of a weak
shock just before the flow enters through the inner sonic point.}
\end{center}
\end{figure}

Even when the explicit cooling is absent, the CENBOL can oscillate (Ryu, Chakrabarti \& Molteni,
1997, hereafter RCM97) in those situations when in the steady state, standing shock solution is absent.
Figure 2 shows the results of a two-dimensional simulation at two different times where we see that the
shock at its extreme locations. The top panel of the Figure 3 shows the time dependence of the shock location.
This oscillation causes variation in the accretion rate, average density, average thermal energy of the disk which 
are also shown in Fig. 3
As pointed out in Ryu et al. (1997), the oscillation is due to absence of steady solutions for the injected
energy and angular momentum of the flow. This type of oscillations should therefore be very common.

In more recent work of the numerical simulation Chakrabarti, Acharyya \& Molteni, (2004)
(hereafter CAM04) showed that the power density spectra could be best reproduced if the symmetry around the
equatorial plane is relaxed and the shocks were also allowed to oscillate not only along the
radial direction, but also along the vertical direction. Normally, in spherically symmetric space-time
these two frequencies are identical, but in axi-symmetric curved space-times (such as Kerr) they may be different.
The simulation in Kerr spacetime is being carried out using a pseudo-Kerr potential 
(Chakrabarti \& Mondal, 2006; Mondal \& Chakrabarti, 2006)
and the results would be reported soon. 

In Fig. 4, the configurations of the disk having both the radial and vertical oscillations of the CENBOL
are shown. In Figs. 5(a-d), we present resulting variation in emitted luminosity with time
(light curves) for four accretion rates increasing from (a) to (d). The mass of the black
hole is chosen to be $10 M_\odot$. In Figs. 6(a-d) the power density spectra (PDS) are shown.
It is clear that the light curves and the PDS have similar characteristics of a typical
$\chi$ class (Belloni et al. 2000) of GRS1915+105. The QPO occurs at frequencies close to break-frequencies. In case (c), the
excessive noise in the light curve resulted in the absence of any prominent QPOs.

As far as the high frequency QPOs are concerned, our understanding is that it could be due to the
oscillation of the inner shock close to the black hole. Figure 7 (upper panel) shows the PDS of one of our
simulations which shows the presence of a high frequency QPO at around $330$Hz. The simulation
results (lower panel) indicate that there is a slight variation near the inner sonic point
at this frequency. For details, see CAM04. 

\begin{figure}
\begin{center}
\vskip 0.0cm
\hskip -10.0cm
\parbox{1.0in}{\epsfig{figure=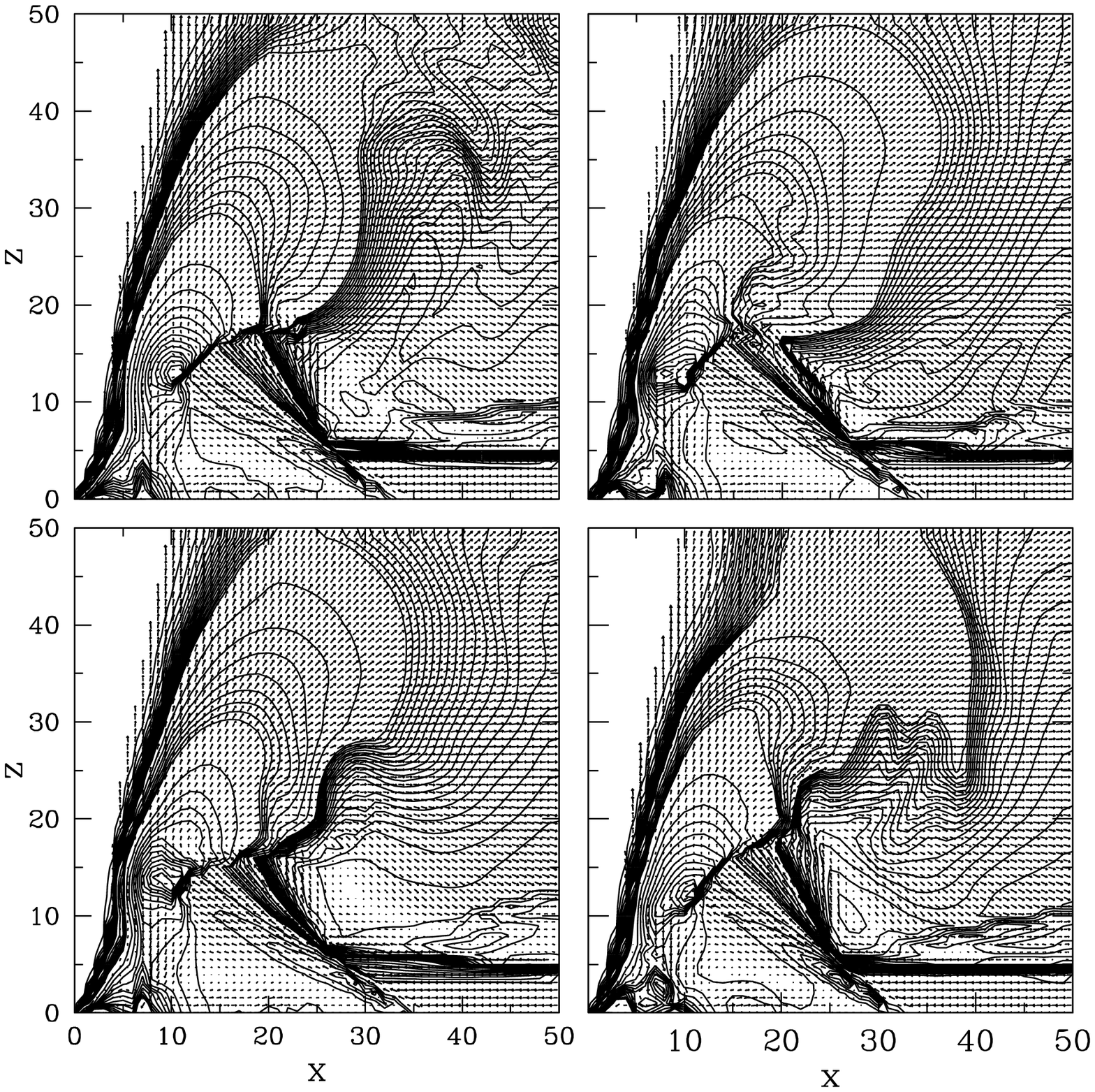,width=5.5in}}
\caption{Contours of constant density are shown for a typical case at four
different times (see text for details) with a shock oscillation causing QPOs. 
Here, ${\cal E} = 0.02$ and $\lambda = 1.75$ have been chosen. Velocity fields 
are superposed to show the turbulence, outflowing jets and the shock. }
\end{center}
\end{figure}

\begin{figure}
\begin{center}
\vskip 0.0cm
\hskip -10.0cm
\parbox{1.0in}{\epsfig{figure=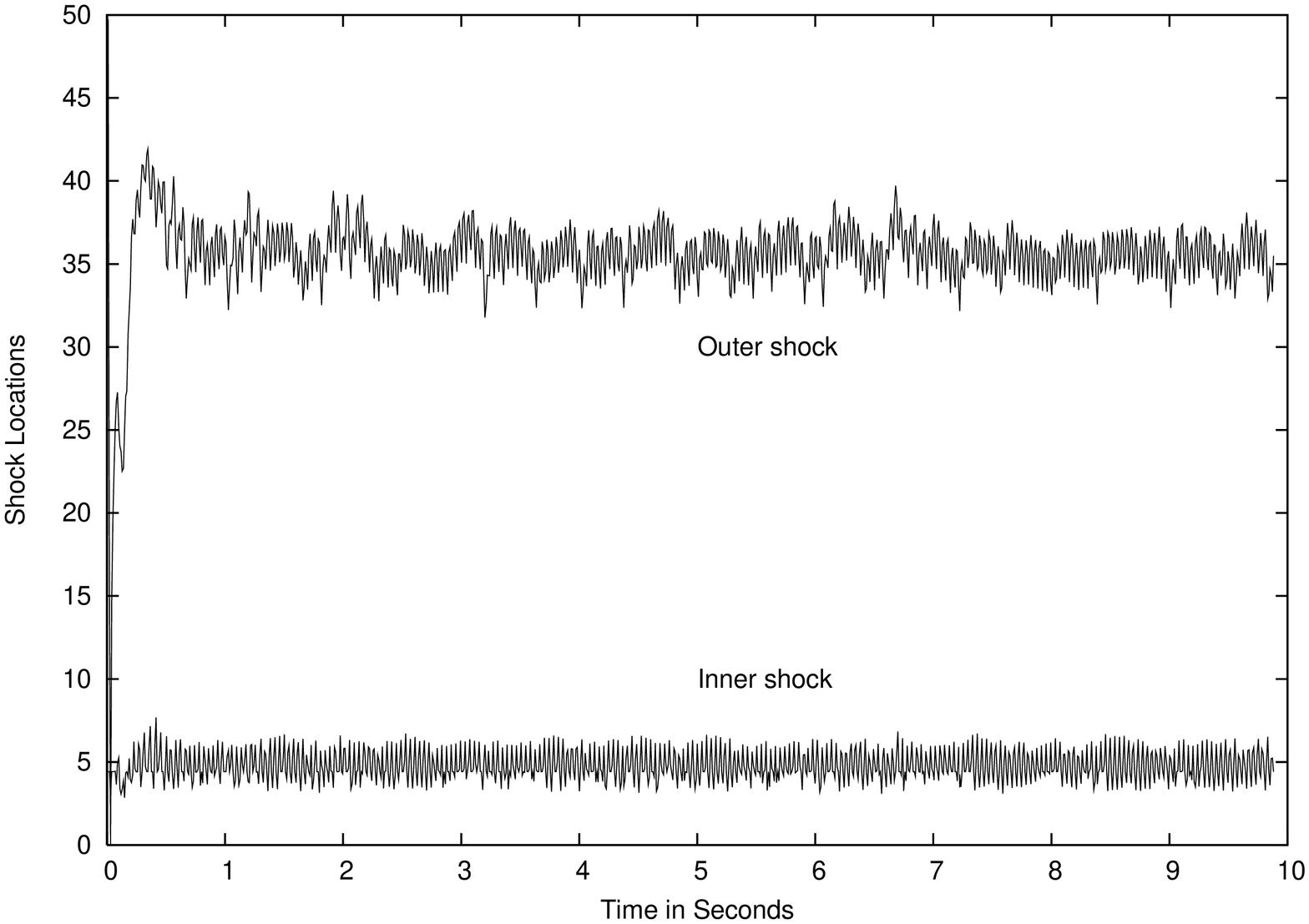,width=3.0in,angle=270.0}}
\figsubcap{a}
\hskip -11.0cm
\parbox{1.0in}{\epsfig{figure=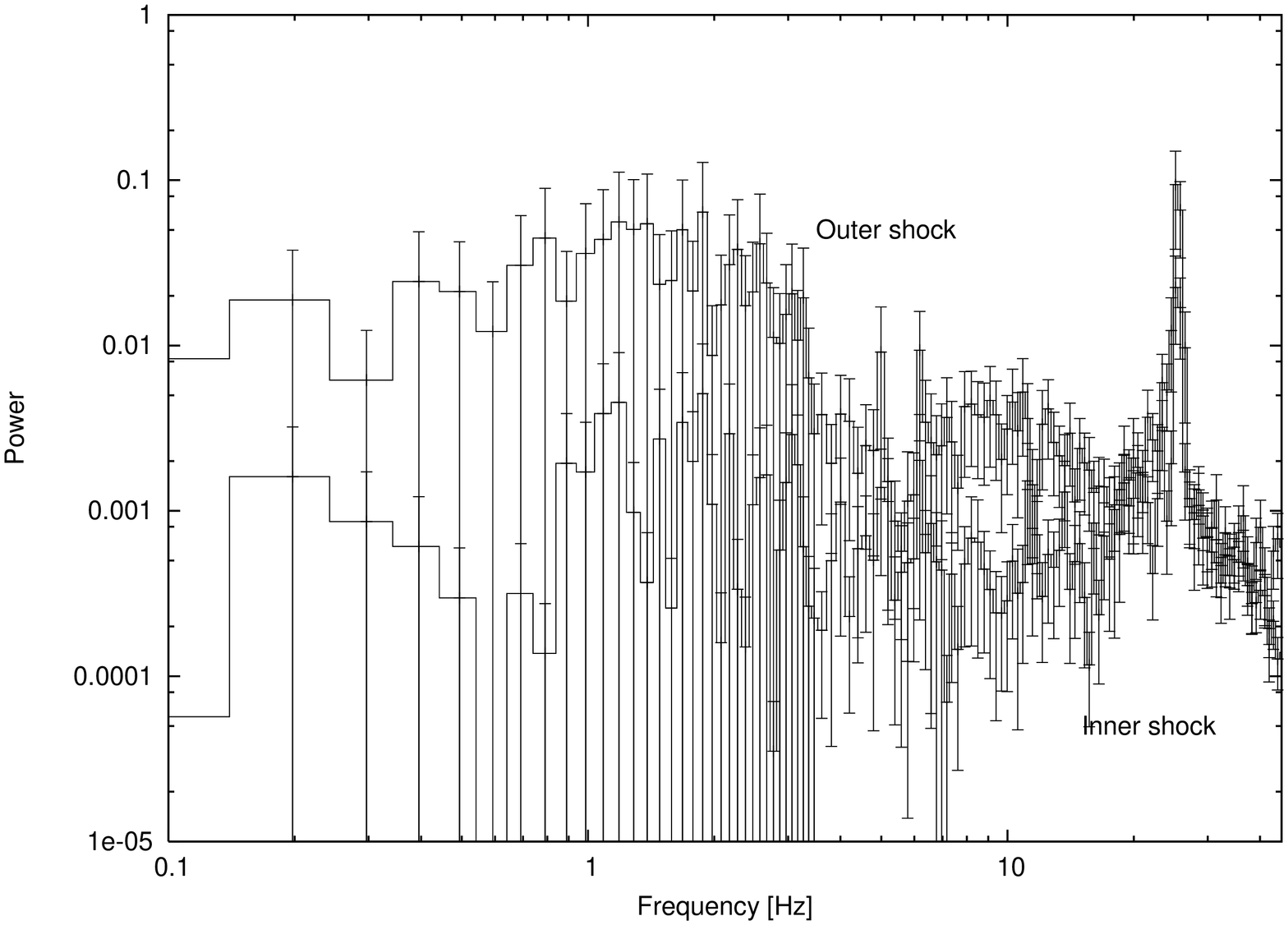,width=3.0in,angle=270.0}}
\figsubcap{b}
\caption{(a) Time variation both the inner and outer shocks and (b) the Fourier
transform (power-density spectrum - PDS) of this variation. Both PDS
show a peak at around $25$Hz. There are broader bumps at lower frequencies as well at which the outer shock
oscillates (Samanta, Chakrabarti \& Ryu, 2007). The simulation parameters are as in Fig. 8 above. }
\end{center}
\end{figure}

\begin{figure}
\begin{center}
\vskip 0.0cm
\hskip -10.0cm
\parbox{1.0in}{\epsfig{figure=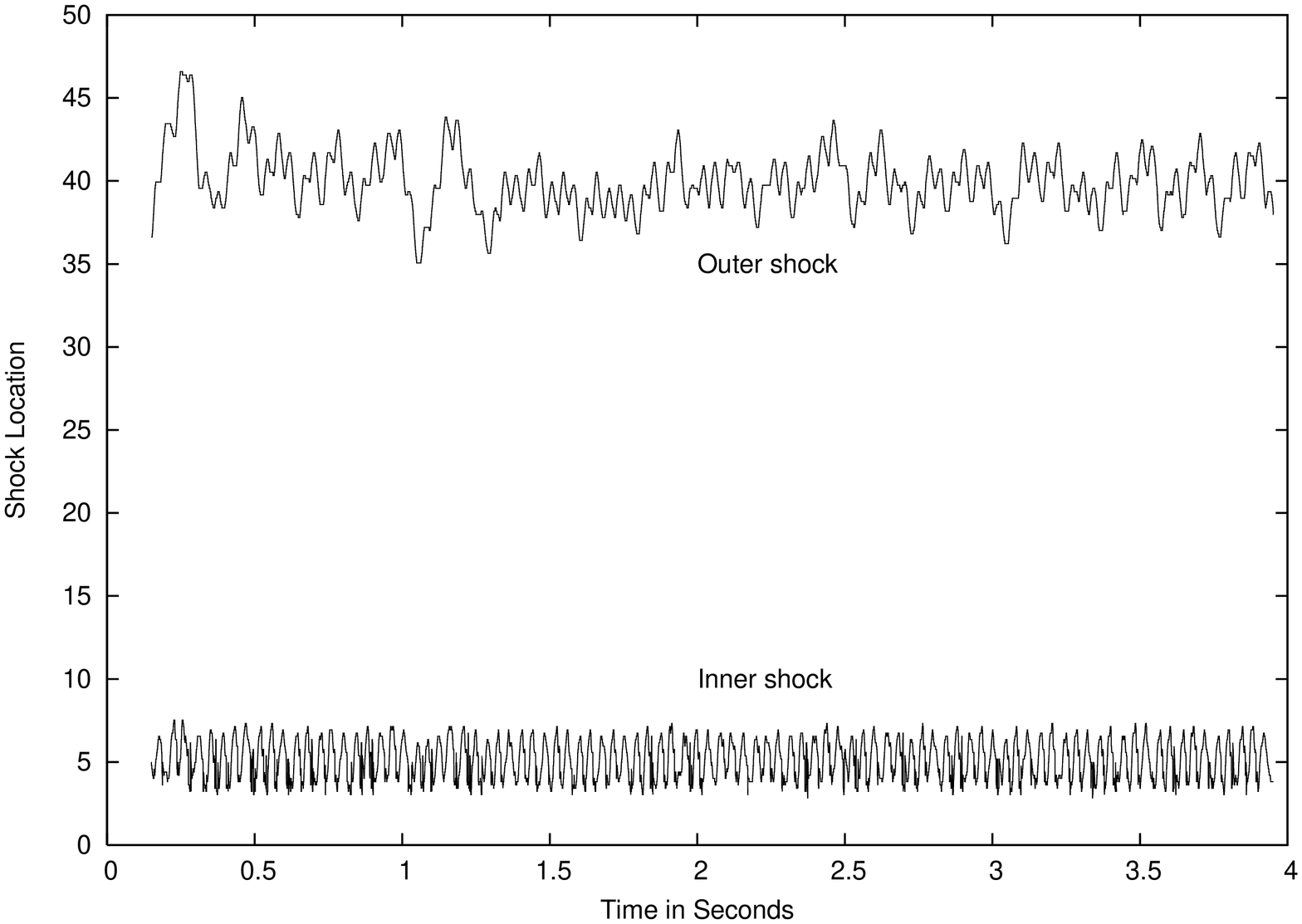,width=3.5in,angle=270}}
\figsubcap{a}
\vskip 0.0cm
\hskip -10.0cm
\parbox{1.0in}{\epsfig{figure=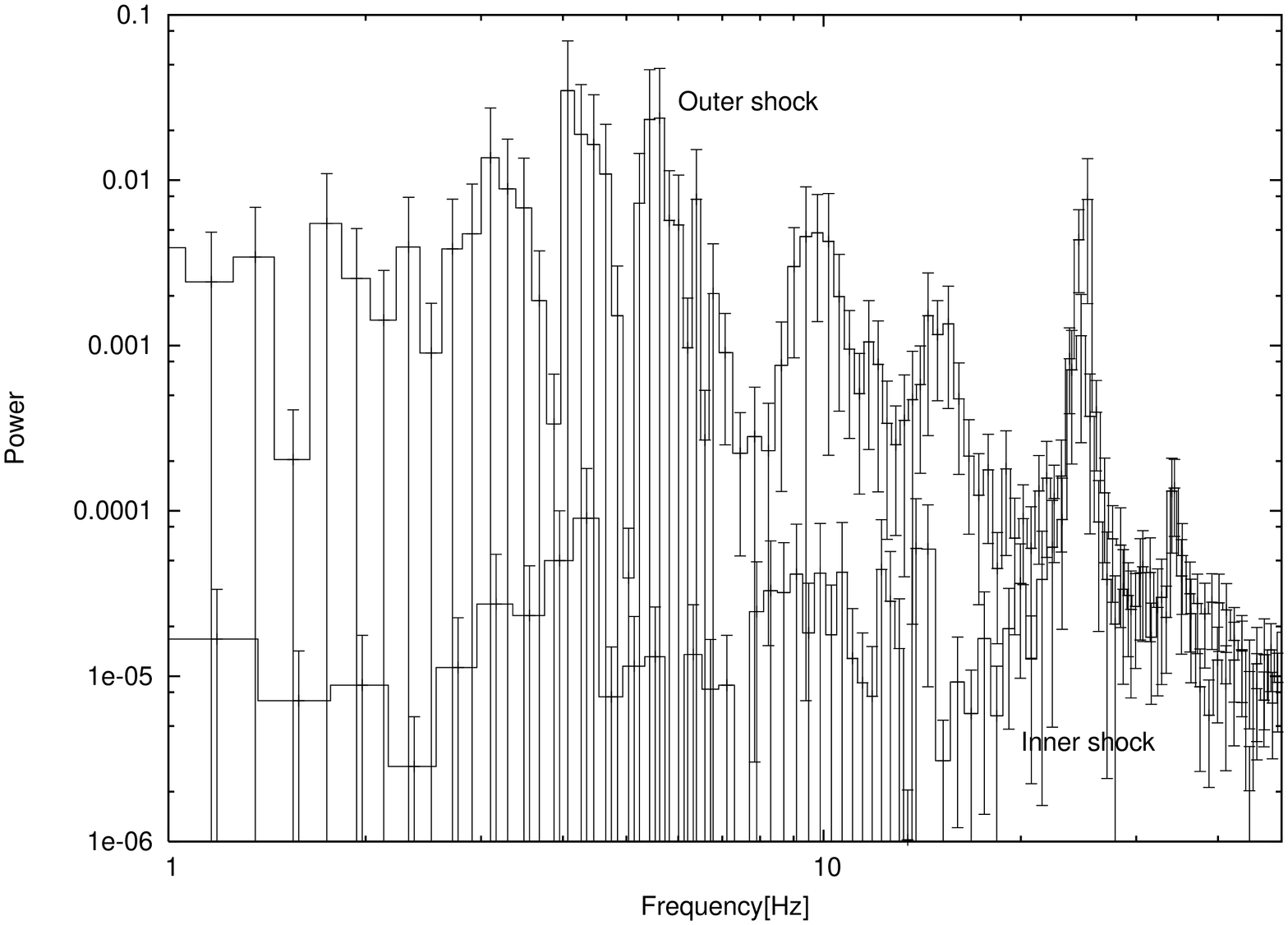,width=3.5in,angle=270}}
\figsubcap{b}
\caption{(a) Time variation of the shock locations for energy ${\cal E}=0.04$ and angular
momentum $\lambda=1.75$; and (b) their Fourier transforms indicating the existance of 2:3 ratio twin 
peaks due to the outer shock oscillation. However, such ratio may be uncommon in axisymmetric shocks 
(Samanta, Chakrabarti \& Ryu 2007).}
\end{center}
\end{figure}

\begin{figure}
\begin{center}
\vskip -4.0cm
\hskip -6.0cm
\parbox{1.0in}{\epsfig{figure=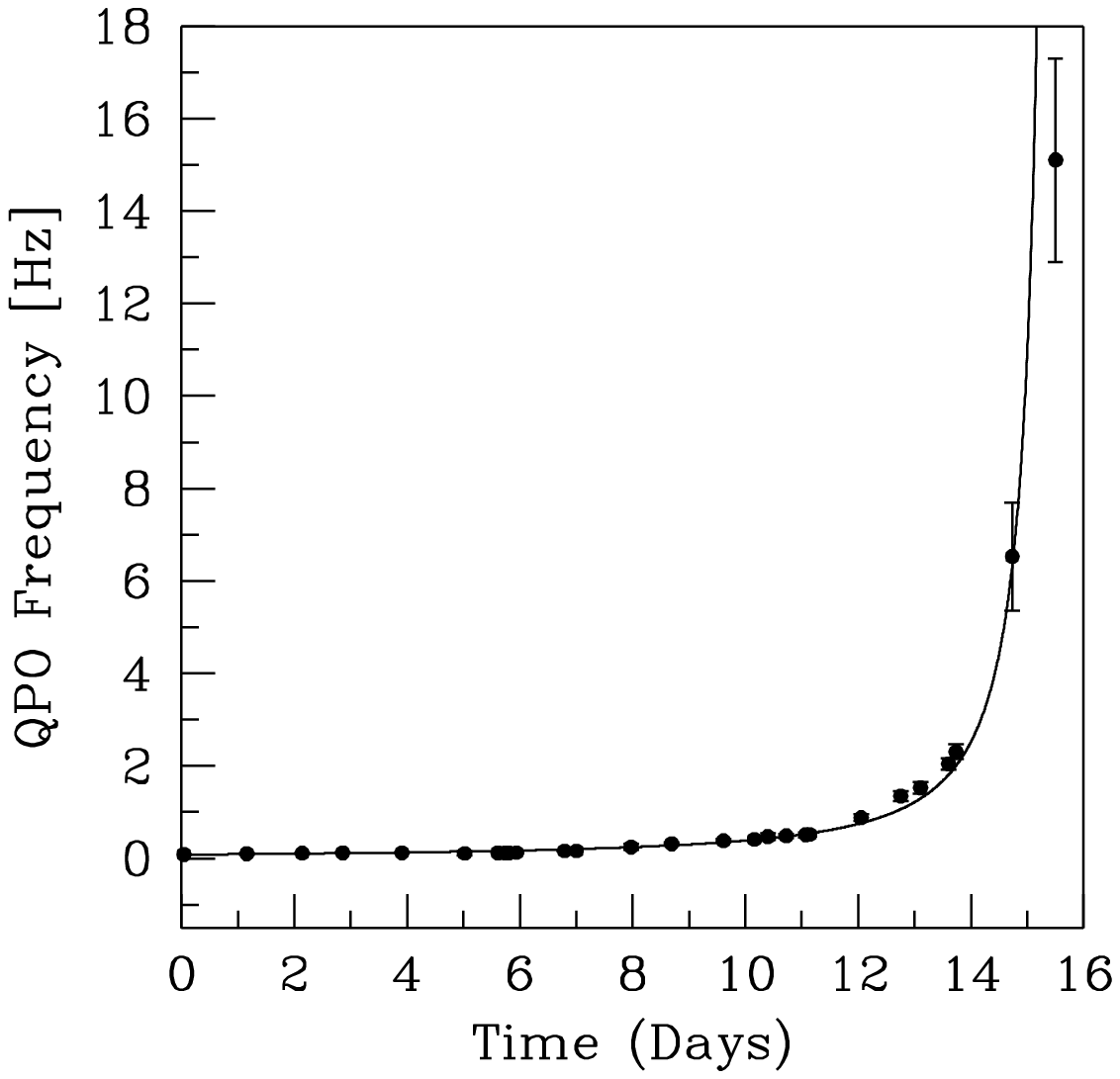,width=4.5in}}
\figsubcap{a}
\vskip 1.0cm
\hskip -10.0cm
\parbox{1.0in}{\epsfig{figure=apo3revfig11b.eps,width=4.5in}}
\figsubcap{b}
\caption{Variation of the QPO frequencies during (a) the onset (Chakrabarti et al. 2005)
and (b) the decline phase of the outburst of GRO1655-40 which started in February, 2005
(from Chakrabarti, Debnath, Nandi \& Pal 2007). Superimposed on them are the analytical curves 
showing signatures of the (a) shock-drift with constant velocity at the (a) onset and 
(b) accelerated shock outwards in disk (first 3.5d in the plot) and in the jet 
(after the first 3.5d).} 
\end{center}
\end{figure}

The results we just reported from SPH simulation are corroborated also even when the
TVD code is used for numerical simulation. Fig. 8 shows a simulation with a typical
parameter set  of ${\cal E}=0.02$ and $\lambda = 1.75$ (Samanta, Chakrabarti \& Ryu, 2007).   
Here, energy is in units of $c^2$ and $\lambda$ is units of $2GM/c$. The density contours (solid curves)
and velocity vector fields (arrows) at four different times (t=1.55s upper left; t=1.60s lower left; 
t=1.65s upper right and t= 1.70s lower right) are shown. Subtle changes in the outer shock 
location ($r \simeq 32$) and inner shock location ($r \simeq 8$) can be seen even 
when the time difference is only a fraction of a second. The oscillations
of shock locations are plotted in Fig. 9a and the Fourier transformation of these locations are plotted 
in Fig. 9b. We clearly see that both shocks are oscillating at around $\nu=25$Hz. Very often we 
see multiple QPOs. In Figs. 10(a-b) we present the time variation of the inner and the outer 
shock locations and their Fourier transforms when ${\cal E}=0.04$ was chosen. 
We clearly see evidence of QPOs at $\nu \sim 24$ and $\nu \sim 35$Hz which is 
close to 2:3 ratio. This phenomenon is not always observed and it is doubtful if the
observed 2:3 ratio is due to axisymmetric shocks at all. The oscillations of the two shocks are
coupled together. This nature of all possible oscillation frequencies needs to be studied
before any firm conclusion about the 2:3 ratio could be reached.

In Table 1, we show a few stellar and super-massive mass black hole candidates which have exhibited QPOs.
In supermassive black holes it is not common to talk about the variabilities 
as QPOs. But we believed that due to the generic nature of the QPOs, the quasi-regular variabilities are
indeed due to shock oscillations and mentioned this even in early to mid ninties
(Chakrabarti \& Wiita, 1992, 1993; Molteni, Sponholz \& Chakrabarti, 1996). 

QPO physics in black hole candidates becomes more transparent when an outburst 
is studied in detail. In February 2005, the GRO 1655-40 exhibited an outburst which
is more exciting than that which occured in 1996. Here, one could track, how on a daily basis
the QPO frequency rises and finally QPO itself disappears suddenly. By fitting the 
QPO frequencies with time it became clear that the oscillating shock drifts in in 
at a constant velocity of nearly 20 meters a second when the out burst started.
Figure 11a shows the frequency change with time and its fit with constant shock-drift
model (Chakrabarti et al. 2005). At the decline phase of the outburst, the shock 
drifts outward at a constant acceleration (Fig. 11b). The changes in the spectral characteristics 
indicate that the decline occurs when the oscillating shock propagates away
in the jets (Chakrabarti et al. 2007). Here the spectrum becomes more and more
dominated by the power-law component of the jets. This drifting in of the accretion shock
and drifting out of the jet shock could be common in many systems.

\noindent Table 1: Black holes with QPO frequencies and other related parameters$^+$.
{\small
\begin {tabular}{|c|c|c|c|c|}
\hline
Name  & $M_{bh}/M_\odot$ & Inclination ($^o$) & HFQPO (Hz) & LFQPO (Hz)\\
\hline
XTE J1550-564 & 8.4-10.8 & 67$\leq$ i $\leq$ 77.4$^{[1]}$ & 92(?), 184, 276 & 0.1-10\\ 
\hline
GRO J1655-40 & 6-6.6 & 70.2$\pm$1.9$^{[2,3]}$ & 300, 450 & 0.1-28\\
\hline
GRS 1915+105 & 10.0-18.0 & 66$\pm$2$^{[4]}$ & 41, 67, 113, 168 & 0.001-10\\
\hline
H 1743-322 & ?? & 60-70$^{[5]}$ & 160, 240$^{[5]}$ & 0.01-22$^{[5]}$\\
\hline
XTE J1650-500 & ?? & 50$\leq$ i $\leq$ 80$^{[6]}$ & 55$\pm$5, 103$\pm$28, & 1.3-10.9$^{[7]}$\\ & & & 139$\pm$8, 
204$\pm$16, & \\ & & & 250$\pm$5, 280$^{[7]}$ & \\
\hline
XTE J1859+226 & 7.6-12.0 & ?? & 82, 150, 190$^{[8]}$; & 0.5-10\\
\hline
Cyg X-1 & 6.8-13.3 & 35 & 134$^[9]$ & 0.035-12\\
\hline
GRO J0422+32 & 3.7-5 & 45$\pm$2$^{[3,10]}$ & - & 0.035-32\\
\hline
GRS 1009-45 & 3.6-4.7: & 67$^{[11,12]}$ & - & 0.04-0.3\\
\hline
XTE J118+480 & 6.5-7.2 & 70$^{[13,14]}$ & - & 0.07-0.15\\
\hline
4U1543-475 & 8.4-10.4 & 20.7$^{[3,15]}$ & - & 7\\
\hline
GS 2000+251 & 7.1-7.8 & 64$\pm$1.3$^{[3,11]}$ & - & 2.4-2.6\\
\hline
MCG-6-30-15$^{++}$ & $\sim  1E+06^{[16]}$ & ?? & - & $1E-04^{[16]}$\\
\hline
RX J0437.4-4711 & $\sim 1E+08^{[17]}$ & ?? & - & $1.3E-05^{[17]}$\\
\hline
NGC 4051 & $1.4 \times 1.E+06^{[18]}$ & 50$^{[18]}$ & - & $\sim 1.2E-07^{[18]}$\\
\hline
Ark 564 & $\sim 8E+06^{[19]}$ & ?? & - & ??\\   
\hline
OJ 287 & 16$\pm1.5E+09^{[20]}$ & ?? & - & $(1-3)E-09^{[20,21]}$\\ 
\hline
M82 X-1 & $\sim 1E+03^{[22]}$, 25-500$^{[23]}$ & ?? & - & (54,58.5,67,87,113,\\ 
& & & & 166)$E-03^{[22-25]}$\\
\hline
NGC 5408 X-1 & 100$^{[26]}$ & ?? & - &  $1E-03^{[27]}$\\
\hline
Holmberg IX X – 1 & 50-200$^{[28]}$ & ?? & - & 202.5 $1E-03^{[28]}$\\
\hline
\end {tabular}}
\noindent $^+$ Generally from McClintock \& Remillard (2006); $^{++}$ QPO frequencies in the massive black holes have been
computed from the time periods given in the paper. \\
\small {$^{[1]}$ Orosz et al., 2002; $^{[2]}$ Greene, Bailyn, \& Orosz, 2001; $^{[3]}$ Garcia et al., 2001; $^{[4]}$
Fender et al., 1999; $^{[5]}$ Homan et al., 2005; $^{[6]}$ Orosz et al., 2004; $^{[7]}$ Homan et al., 2003; 
$^{[8]}$ Cui et al. 2000; $^{[9]}$ Remillard et al., 2006; $^{{[10]}}$ Gelino \& Harrison 2003; $^{{[11]}}$ 
Orosz, 2002; $^{{[12]}}$ Hameury et al. 2003; $^{{[13]}}$ Orosz et al.  2004; $^{{[14]}}$ McClintock et al. 2003;
$^{{[15]}}$ Orosz  et al. 2004; $^{{[16]}}$ Arevalo, 2005; $^{{[17]}}$ Halpern \& Marshall, 1996; $^{{[18]}}$ 
Peterson et al., 2000; $^{{[19]}}$ Shemmer et al. 2001; $^{{[20]}}$ Valtonen et al., 2006a,; $^{{[21]}}$ Valtonen et al., 2006b;
$^{{[22]}}$ Fiorito \& Titarchuk, 2004; $^{{[23]}}$ Dewangan, Titarchuk \& Griffiths, 2006; $^{{[24]}}$
Strohmayer \& Mushotzky, 2003; $^{{[25]}}$ Mucciarelli et al. 2006; $^{{[26]}}$
Soria 2004; $^{{[27]}}$ Strohmayer et al. 2007; $^{{[28]}}$ Dewangan, Griffiths \& Rao, 2006}

\section{QPO Frequencies Having 2:3 Ratio}

Several black hole candidates, such as GRS 1915+105, GROJ 1655-40, H 1743-322 and XTE 1550-564 exhibit twin
high frequency quasi-periodic oscillations (QPOs) whose ratio is 2:3.
This behaviour seems to occur only for high frequencies (though in GRS 1915+105, 41Hz and 67 Hz, with a ratio
of almost 2:3 is not of that high frequency; this only indicates that GRS 1915+105 is of higher mass than 
other average stellar mass black holes),  
i.e., when the oscillating physical region is located closer to the black hole. Second, the objects
which exhibit them have the disks inclined at an angle of about $\sim 60-80$ degrees or more. Third
and perhaps more important, QPOs of this ratio are seen when the spectral state is becoming softer. Moreover,
the fundamental ferquency is not seen at all (but see, Remillard et al. 2002 where it is claimed that
the fundamental is present in  XTE J1550-564 and GRO J1655-40). Here we argue that 
this 2:3 ratio is actually due to non-axisymmetric spiral shocks in the disk near the black hole.

\section{Why Spiral Structures Should Form In The Disk}

Spiral structures are common in accretion flows since any density enhancements would be sheared before it falls 
on to the black hole. However, to have sustained structures one may require sustained perturbation such 
as a binary companion. In a strictly two dimensional flow, the turbulent energy is known to be concentrated in
larger scales and in strictly three dimension, the energy cascades into smaller and smaller scales
(e.g., Frisch, 1996). However, an accretion disk is neither very thin (two dimensional) nor very thick 
(three dimensional), especially when spectrally soft states are achieved and the CENBOL size (both 
vertically and radially) decreases due to the fall of pressure. Hence the turbulent energy would neither cascade 
in to the largest scale (single armed spiral), nor to the smallest scale (total chaos), but
perhaps to intermediate scales (two or three arms). Numerical simulations indicate that the tidal processes 
due to a binary companion do not induce one armed spiral shocks. Simulations of 3D flows also show that the one 
arm is produced only due to ram pressure induced by eccentric orbits (Hayasaki \& Okazaki, 2005).
Single armed spirals were seen to grow rapidly in a thin, two dimensional, constant angular momentum
simulation (Blaes \& Hawley, 1988) but as the angular momentum distribution is changed towards
Keplerian distribution, the growth rate is reduced. Single armed spiral waves may exist only 
in nearly Keplerian disks provided only a single central source of perturber is present 
(Lee \& Goodman, 2005). In theoretical studies of self-similar spiral shocks it was found (Chakrabarti 1990)
that when the adiabatic index $\gamma$ is closer to $1.67$, the density waves become fragmentary and no 
large scale structures can form. As $\gamma$ starts going down to $1.5$ or less, the radiation pressure 
starts dominating and the standing shocks may form since the flow starts having more
than one saddle type sonic points. Steady state spirals may also form.
However when $\gamma$ goes down even further to $1.2$ 
or so, and the gas becomes cooler. Thus, as the flow goes to spectrally soft
state, the spirals may fragment into two to three pieces but not to one. All these
results suggest that the formation of more than one arm is more probable
when we have a realistic disk. At a given point in time, how many arms would form
will depend on the nature of the viscous processes which dictate the
angular momentum distribution inside the disk.

Because the shape of the inner disk becomes increasingly thiner when the
spectrally softer states are reached (due to collapse of the CENBOL), we believe
that the 2:3 ratio observed in QPO frequencies in
several objects is directly related to the behaviour 
of turbulence in a `2.5-dimensional' accretion disk which is neither thick nor thin
(Chakrabarti et al. 2006).
As the accretion rate increases and the axisymmetric shock is
pushed inside due to ram pressure, it breaks partially and causes the spiral structures to form
(which may be sustained due to tidal effect of the companion). Thus, these high frequency QPOs
at 2:3 ratio occur when the spectrum is becoming softer.

\section{Fragmentation of Spiral CENBOL Close to a Black Hole}

\begin{figure}
\vbox{
\vskip 0.0cm
\hskip 0.0cm
\centerline{\psfig{figure=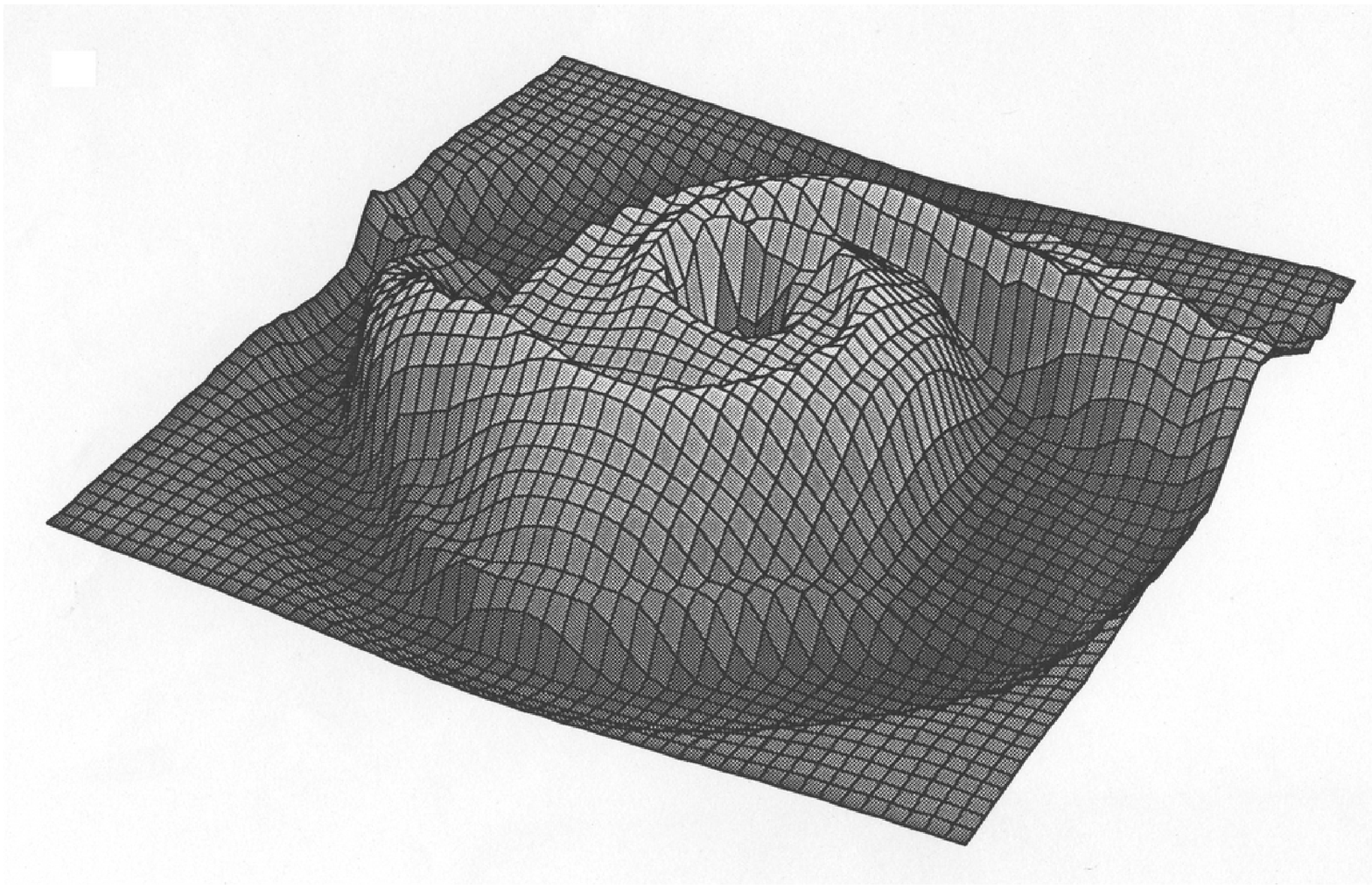,height=7truecm,width=7truecm,angle=90}}
\centerline{\psfig{figure=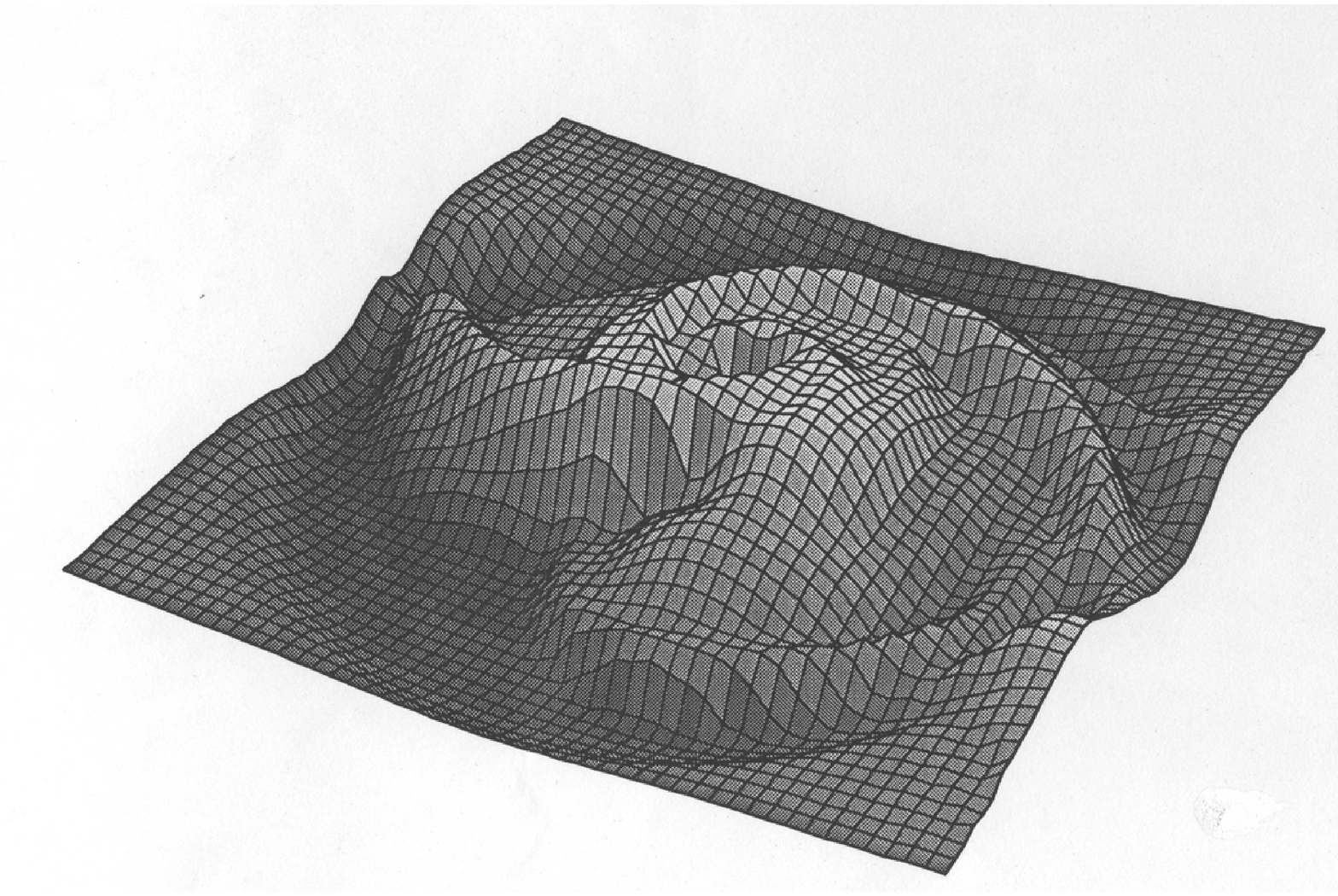,height=7truecm,width=7truecm,angle=90}}
\centerline{\psfig{figure=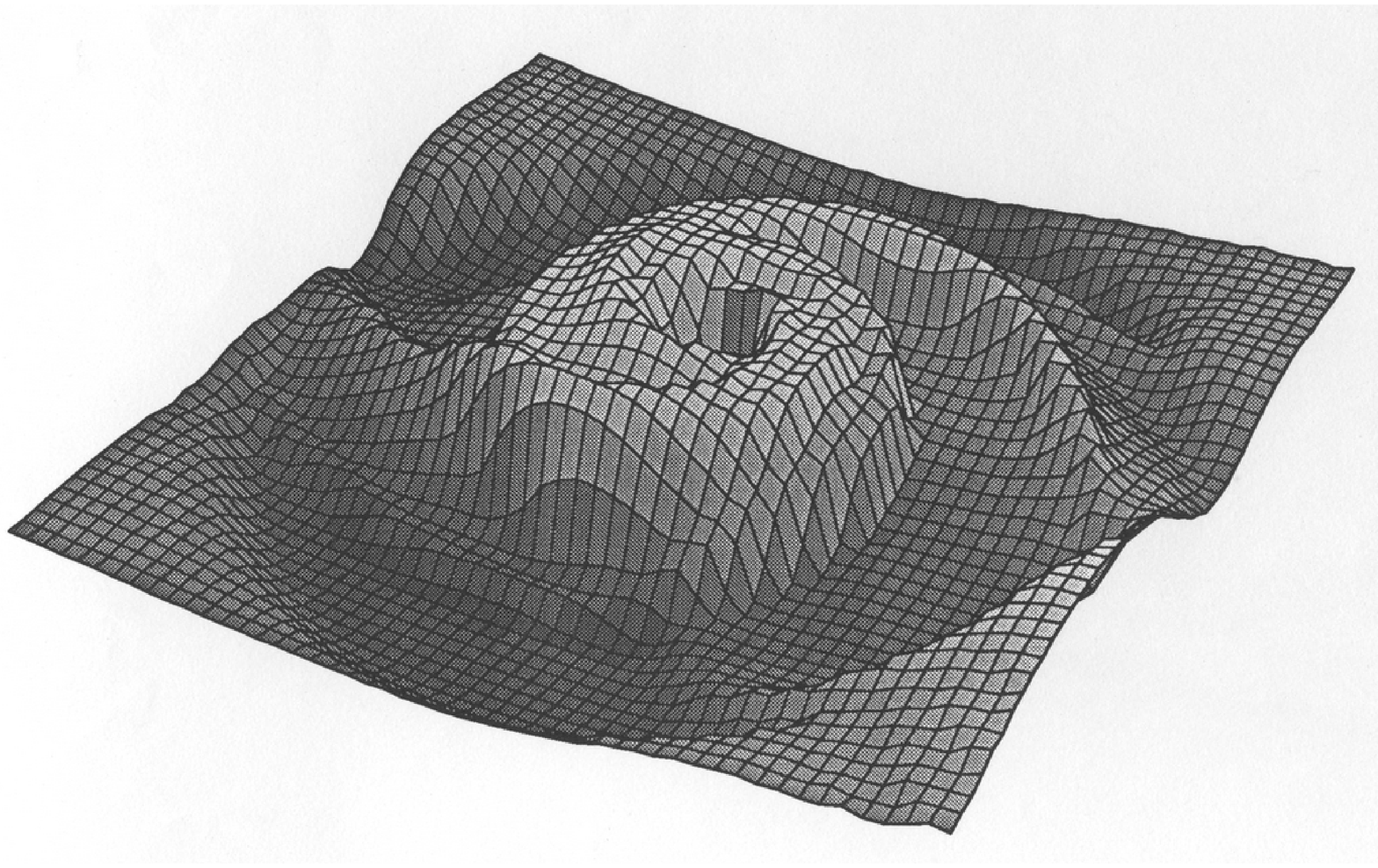,height=7truecm,width=7truecm,angle=90}}
}
\vspace{-0.5cm}
\caption[]
{Variation of the shape of the spiral CENBOL in an accretion flow
in presence of a binary companion. Two armed spiral in (a) is changed to three
armed spiral in (b) before going back to a two-armed spiral again. The curves are
drawn at $T=2.296, 3.422$ and $3.746$ respectively, where time is in units of the
orbital period. The adiabatic index $\gamma=1.2$ was used and the companion
of equal mass has been used in the simulation (Chakrabarti, 1996; Chakrabarti et al. 2006)}
\end{figure}

\begin{figure}
\vbox{
\vskip 0.0cm
\hskip 0.0cm
\centerline{ \psfig{figure=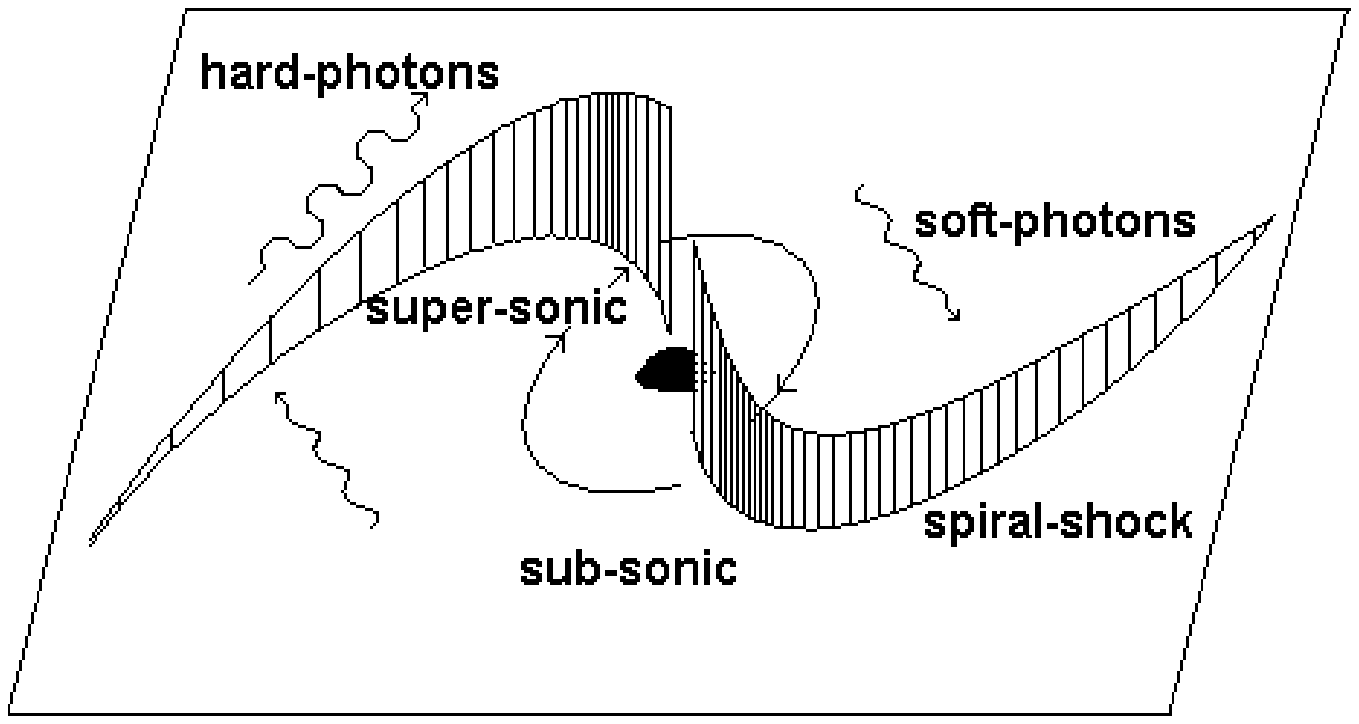,height=8truecm,width=8truecm}}
\figsubcap {a}
\centerline{ \psfig{figure=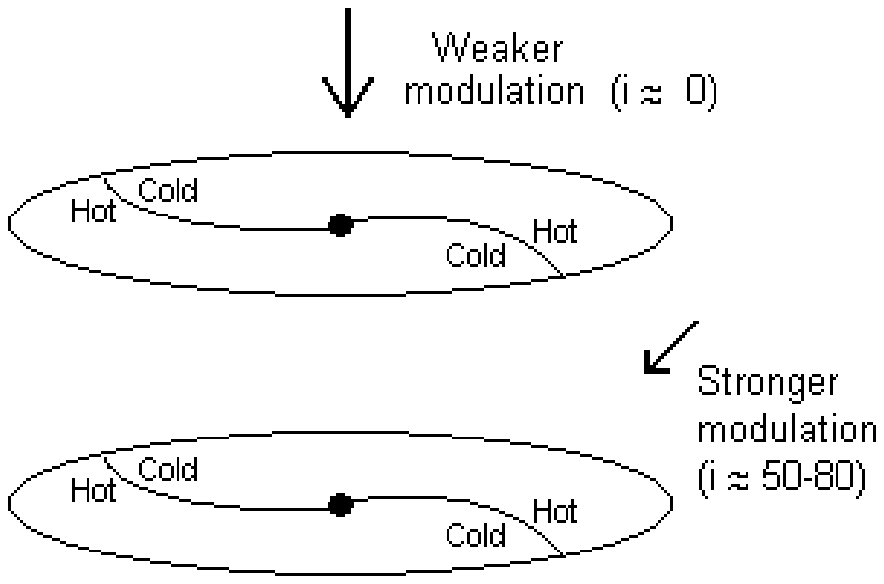,height=8truecm,width=8truecm}}
\figsubcap {b}}
\vspace{0.0cm}
\caption[]{Non-axisymmetric CENBOLs puffed up to intercept
soft photons from the pre-shock flow and re-radiate them in hard X-rays. Oscillations
of these regions cause higher frequency QPOs. When the pattern changes to a three-armed
spiral, QPO frequency increases further by another fifty percent, causing the
appearance of QPOs at 2:3 ratio (Chakrabarti et al. 2006). (b) The modulation increases
when the inclination angle is high due to shadowing effect.}
\end{figure}

We present results of numerical simulations in presence of a binary companion in an adiabatic
accretion disk. Figures 12(a-c) show the density (Z-axis) distribution as a function of X and Y
coordinates. The simulation was carried out with $\gamma=1.2$ and $q=M_x/M_c =1$, where,
$M_x$ and $M_c$ are the masses of the compact object and the companion respectively. The three
figures are drawn at $T=2.296$, $3.422$ and $3.746$ respectively (Chakrabarti, 1996).
Here, time is measured in units such that 2 pi is the orbital period. Thus the number
of arms are changing from (a) two to (b) three to
(c) two respectively in about 0.5-1 orbital period. The post-shock region (which may be called
Non-Axisymmetric CENtrifugal pressure supported BOundary Layer, or NACENBOL) will continue
to be hot and will continue to Comptonize soft photons as the axisymmetric CENBOLs do. 
In order that the effect of non-axisymmetry is reflected in the observed
lightcurve, the inclination angle must be large, otherwise modulation due to non-axisymmetry
will be absent or weak. The strongest emitting region from these shocks are very
close to the inner region of the disk (Fig. 13a) and when compared with the procedure
by which CENBOL oscillates (Chakrabarti, Acharyya \& Molteni, 2004) we expect that the
NACENBOLs would also oscillate around the mean location at time scales comparable to the dynamical time
(time to traverse from one shock to the other) at the inner region. Thus, the ratio of the
QPOs would be the ratio of the dynamical time of the flow having two and three shocks. As schematically
shown in Fig. 13b, the modulation is the highest when viewed edge on due to shadowing effect.

\section{Concluding Remarks}

The advective disks which include oscillating axi-symmetric and non-axisymmetric 
shocks reproduce both the low frequency and high frequency QPOs 
very well. The fact that QPO frequencies change
with luminosity may be due to the variation of cooling time scale with accretion rate.
Unlike other models of QPOs which resort to various types of vibrations, our solution
shows that (a) the low frequency QPOs should generally occur during the transition to the 
low-hard state or during a low-hard state. Very low frequency QPOs are associated with 
the filling time of the sonic radius. (b) Since the jets are associated with 
CENBOL, it is expected that when QPOs are seen, there should be jets/outflow
activities as well. (c) Since QPOs are supposed to be due to the oscillations of the
CENBOL which shrinks with the increase of the cooling rate, i.e., accretion rate, the
frequency of oscillation should generally rise with the accretion rate. This will be contradicted
if, say, viscosity goes down and the angular momentum goes up which increases the size of the CENBOL.

We showed a curious simulation result (Samanta, Chakrabarti \& Ryu, 2007) that very often 
there are two shocks in the flow: one is due to the usual centrifugal barrier and the 
other is due to the turbulence and both are coupled together in such a way the high 
frequency oscillation of the inner shock is also iduced to the outer shock. But the outer 
shock also has lower, very often non-harmonic, components. This study may be valuable in 
understanding multi-frequency QPOs in black holes.

Occasionally, QPOs at high frequencies are observed at $\nu_2 /\nu_3= 2/3$ ratio. 
What we expect is that when the CENBOl is cooled down to the extent that the 
axisymmetric shock moves closer, some kind of non-axisymmetric perturbation 
causes it to break in to two or three shocks alternatively. Thus the fundamental
(due to axisymmetric shock) disappears and harmonics set in, typically one 
becomes more intense than the other. Our simulations show this alternate break up
in order of fraction of an orbital period also explains why objects with disks 
having high inclination angle are more favorable to have QPOs with such a special 
frequency ratio. However, intrinsic variation of the shape of the NACENBOL causes
differential variation of the intercepted soft photons. Thus, emitted hard photon 
numbers should continue to be modulated even at a smaller inclination angle also. 
But when viewed nearly end on, the modulation would be insignificant in comparison
to the background photon.

\noindent{Acknowledgments:}

SKC acknowledges kind hospitality of ICRA, Pescara where this work was partly completed.
Mr. H. Ghosh and Mr. P.S. Pal  acknowledge the support of the ISRO RESPOND project. D. Debnath acknowledge
the CSIR-NET fellowship and R. Sarkar acknowledge the support of ASTROSAT JRF position.


\begin{thebibliography}{}

\bibitem{} Arevalo, P. et al., 2005, A\&A, 430, 435
\bibitem{} Belloni, T. et al., 2000, A\&A, 355, 271
\bibitem{} Blaes, O.M. \& Hawley, J.F., 1988, {\it Astrophys. J.},  326, 277
\bibitem{} Chakrabarti, S.K., 1989, ApJ, 347, 365
\bibitem{} Chakrabarti S. K., 1990a, Theory of Transonic Astrophysical Flows, World Scientific Publishing, Singapore.
\bibitem{} Chakrabarti S. K., 1990b, MNRAS, 243, 610
\bibitem{} Chakrabarti, S. K., 1996,  Phys. Rep., 266, No. 5 \& 6, 229
\bibitem{} Chakrabarti, S.K., 1999, A \& A, 351, 185
\bibitem{} Chakrabarti, S.K., Acharyya, K. \& Molteni, D., 2004, Astron. Astrophys., 421, 1
\bibitem{} Chakrabarti, S.K. \& Das, S., 2004, MNRAS, 349, 649
\bibitem{} Chakrabarti S. K. \& Manickam S. G., 2000, ApJ, 531, 41
\bibitem{} Chakrabarti, S.K. \& Mondal, S.K., 2006, MNRAS, 369, 976
\bibitem{} Chakrabarti, S. K. \& Molteni D., 1993, ApJ, 417, 671.
\bibitem{} Chakrabarti, S. K. \& Nandi, A., 2000, Ind. J. Phys., 75(B), 1 (astro-ph/0012526)
\bibitem{} Chakrabarti, S.K. \& Wiita P.J., 1992, ApJ, 387, 21
\bibitem{} Chakrabarti, S.K. \& Wiita, P.J., 1993, ApJ, 411, 602
\bibitem{} Chakrabarti, S. K., Nandi, A., Debnath, D., Sarkar, R. and Dutta, B.G., 2006, in Proceedings
of Science -- Proceedings of the VIth Microquasar Workshop: Microquasars and Beyond, p. 103
\bibitem{} Chakrabarti S. K., Acharyya K. \& Molteni D., 2004, A\&A, 421, 1
\bibitem{} Chakrabarti, S.K., Nandi, A., Debnath, D., Sarkar, R. \& Datta, B.G., 2005, Ind. J. Phys., 78B, 1
\bibitem{} Chakrabarti, S.K., Debnath, D., Nandi, A. \& Pal, P.S., 2007, ApJ Letters (submitted)
\bibitem{} Cui, W. et al., 2000, ApJ, 535, 123
\bibitem{} Dewangan, G.C., Titarchuk, L.G. \& Griffiths, R.E., 2006, ApJ, 637, 22
\bibitem{} Dewangan, G.C., Griffiths, R.E., \& Rao, A.R., 2006, ApJ, 641, 125
\bibitem{} Esin, A.A., McClintock, J.E., Narayan, R., 1998, ApJ, 500, 523
\bibitem{} Fender et al., 1999, MNRAS, 304, 865
\bibitem{} Fiorito, R. \& Titarchuk, L.G., 2004, ApJ, 614, 113
\bibitem{} Frisch, U., 1996, Turbulence (Cambridge Univ. Press)
\bibitem{} Halpern, J.P. \& Marshall, H.L., 1996, ApJ, 464, 760
\bibitem{} Hameury, J.M., Barret, D., Lasota, J.P., McClintock, J. E., Menou, K., Motch, C., Olive, J.F., \&
\bibitem{} Hayasaki, K. \& Okazaki, T., 2005, {M.N.R.A.S}, 2005, 360 L15
\bibitem{} Homan et al., 2003, ApJ 586, 1262
\bibitem{} Homan et al., 2005, ApJ, 623, 383
\bibitem{} Lanzafame, G., Molteni, D. \& Chakrabarti, S.K., 1998, MNRAS, 299, 799
\bibitem{} Lee, E. \& Goodman, J. 2005,  { M.N.R.A.S.}, 308, 984
\bibitem{} Garcia, M. R., McClintock, J. E., Narayan, R., Callanan, P., Barret, D., \& Murray, S. S., 2001, ApJ, 553, L47
\bibitem{} Gelino, D. M. \& Harrison, T. E., 2003, ApJ, 599, 1254
\bibitem{} Greene, J., Bailyn, C. D., \& Orosz, J. A. 2001, ApJ, 554, 1290
\bibitem{} McClintock, J. E., Narayan, R., Garcia, M. R., Orosz, J. A., Remillard, R. A., \& Murray, S. S.,
2003, ApJ, 593, 435
\bibitem{} McClintock, J.E. \& Remillard, R.A., 2006, “Compact Stellar X-Ray Sources”, (Eds.) W H.G. Lewin \& M. van der Klis, Cambridge Univ. Press, 157
\bibitem{} Molteni D., Lanzafame G. \& Chakrabarti S K., 1994, ApJ, 425, 161
\bibitem{} Molteni D., Ryu D. \&  Chakrabarti S. K., 1996, ApJ, 470, 460
\bibitem{} Molteni D., Sponholz H. \& Chakrabarti S. K., 1996, ApJ, 457, 805
\bibitem{} Mondal, S. \& Chakrabarti S. K., 2006, MNRAS, 371, 1418
\bibitem{} Mucciarelli, P., Casella, P., Belloni, T., Zampieri, L. \& Ranalli, P., 2006, MNRAS 365, 1123
\bibitem{} Orosz, J. A. 2002, in IAU Symp. 212, A Massive Star Odyssey: From Main Sequence to Supernova,
Ed. K. A. van der Hucht, A. Herrero, \& C. Esteban (San Francisco: ASP), 365
\bibitem{} Orosz, J. A. et al., 2002, ApJ, 568, 845
\bibitem{} Orosz, J. A. et al., 2004, ApJ 616, 376
\bibitem{} Orosz, J. A., McClintock, J. E., Remillard, R. A., \& Corbel, S., 2004, ApJ, 616, 376
\bibitem{} Peterson, B. et al., 2000, ApJ, 542, 161
\bibitem{} Remillard, R. A., Morgan, E. H., McClintock, J. E., Bailyn, C. D. \&  Orosz, J. A.,  1999, ApJ, 522, 397
\bibitem{} Remillard, R. A., Muno, M. P., McClintock, J. E., Orosz, J. A., 2002, ApJ, 580, 1030
\bibitem{} Remillard et al., 2006, American Astronomical Society, HEAD meeting \#9, \#1.35
\bibitem{} Ryu D, Chakrabarti S. K. \& Molteni D., 1997, ApJ, 474, 378
\bibitem{} Samanta, M.M., Chakrabarti, S.K. \& Ryu, D., 2007,  MNRAS (submitted)
\bibitem{} Shakura, N.I. \& Sunyaev, R.A., 1973, A \& A, 24, 337
\bibitem{} Shemmer, O. et al. 2001, ApJ, 561, 162
\bibitem{} Soria, R., 2004, A \& A, 423, 955, 963
\bibitem{} Strohmayer, T.E. \& Mushotzky, R.F., 2003, ApJ, 586, 61
\bibitem{} Strohmayer, T.E., Mushotzky, R.F., Winter, L., Soria, R., Uttley, P. \& Cropper, M. 2007, ApJ, 660, 580
\bibitem{} Valtonen, M. J. et al., 2006a, ApJ, 643, 9
\bibitem{} Valtonen, M. J. et al., 2006b, ApJ, 646, 36
\bibitem{} Webb, N., 2003, A \& A, 399, 631

\end{thebibliography}
\end{document}